# Relationship between Local Molecular Field Theory and Density Functional Theory for non-uniform liquids

A.J Archer[1,a)] and R.Evans[2,b)]

[1]*Department of Mathematical Sciences, Loughborough University, Loughborough LE11 3TU, UK*
[2]*H H Wills Physics Laboratory, University of Bristol, Bristol BS8 1TL, UK*

The Local Molecular Field Theory (LMF) developed by Weeks and co-workers has proved successful for treating the structure and thermodynamics of a variety of non-uniform liquids. By reformulating LMF in terms of one-body direct correlation functions we recast the theory in the framework of classical Density Functional Theory (DFT). We show that the general LMF equation for the effective reference potential $\phi_R(\mathbf{r})$ follows directly from the standard mean-field DFT treatment of attractive interatomic forces. Using an accurate (Fundamental Measures) DFT for the non-uniform hard-sphere reference fluid we determine $\phi_R(\mathbf{r})$ for a hard-core Yukawa liquid adsorbed at a planar hard wall. In the approach to bulk liquid-gas coexistence we find the effective potentials exhibit rich structure that can include damped oscillations at large distances from the wall as well as the repulsive hump near the wall required to generate the low density 'gas' layer characteristic of complete drying. We argue that it would be difficult to obtain the same level of detail from other (non DFT based) implementations of LMF. LMF emphasizes the importance of making an intelligent division of the interatomic pair potential of the full system into a reference part and a remainder that can be treated in mean-field approximation. We investigate different divisions for an exactly solvable one-dimensional model where the pair potential has a hard-core plus a linear attractive tail. Results for the structure factor and the equation of state of the uniform fluid show that including a significant portion of the attraction in the reference system can be much more accurate than treating the full attractive tail in mean-field approximation. We discuss further aspects of the relationship between LMF and DFT.

a) Electronic mail: A.J.Archer@lboro.ac.uk   b) Electronic mail: bob.evans@bristol.ac.uk

## I. INTRODUCTION

Developing accurate theories for the structure, thermodynamics and phase behaviour of non-uniform fluids continues to pose considerable challenges even for a simple model system such as the Lennard-Jones (LJ) fluid. It is well-known that theories must account for the oscillations in density profiles that reflect short ranged correlations, arising from repulsive forces (packing) between atoms, as well as for longer ranged and more slowly varying correlations that are associated with attractive forces. For example, a proper description of wetting or drying transitions at a planar substrate and of capillary condensation or evaporation in a confined fluid requires a quantitatively reliable theory for both types of force. Modern density functional theories (DFT) provide a very accurate description of the short ranged oscillatory structure and the surface tension for hard-particle models and in particular for hard-spheres; for recent reviews see Ref. [1-3]. However, with few exceptions, DFT treatments of adsorption and related phenomena incorporate attractive interactions in a crude mean-field fashion that neglects correlation[3,4]. Modern integral equation theories for inhomogeneous fluids provide an alternative approach. These theories often treat both types of force on equal footing but their implementation is numerically demanding and results are often sensitive to the choice of closure approximation. Moreover, unlike DFT approximations, integral equation theories usually suffer from problems of thermodynamic consistency, important for phase transition studies. A third approach was developed by Weeks, Katsov, Vollmayr and co-workers originally for a LJ fluid[5-9]. More recent extensions of this Local Molecular Field Theory (LMF) have proved remarkably successful in applications to both uniform and non-uniform charged fluids[10-12], for electrostatics in models of confined water[13-15] and for uniform polar molecular liquids[16].

The basis of LMF is the idea that there exists a mapping from the full system with pair potential $w(r)$ and external, one-body potential $\phi(\mathbf{r})$ to a mimic system that is described by a (short-ranged) pair interaction $u_0(r)$ and an effective or restructured external potential $\phi_R(\mathbf{r})$. The mimic (reference) system, denoted by subscript R, and the actual system are connected by requiring

$$\rho^{(1)}(\mathbf{r};[\phi]) = \rho_R^{(1)}(\mathbf{r};[\phi_R]) \qquad (1)$$

i.e. the average one-body density in the full system should be the same as that in system R. $\phi_R(\mathbf{r})$ is to be determined self-consistently.

Of course the usefulness of the LMF mapping depends on constructing a suitable reference system and then making a well-chosen (mean-field) approximation to treat the remaining (longer ranged) part of the pair potential $u_1(r)$ defined by

$$w(r) = u_0(r) + u_1(r). \qquad (2)$$

The requirements for choosing the reference system are laid out carefully in Sec. II of the paper by Rodgers and Weeks[14] whose presentation we follow below.

At first sight LMF would appear to have a different basis from DFT in which one writes down an explicit approximation for the intrinsic Helmholtz free energy functional and minimizes the corresponding grand potential functional to obtain the equilibrium density profile and the thermodynamic functions [3, 4, 17]. In the present paper we re-formulate LMF in terms of one-body direct correlation functions and show that the general LMF equation [6, 14] for the effective reference potential $\phi_R(\mathbf{r})$ follows directly from the standard mean-field DFT treatment of attractive forces. We believe that our derivation provides new insight into the relationship between the LMF approximation and DFT treatments.

As mentioned already, accounting quantitatively for wetting and drying transitions is an important goal for theories. It is well-known [18-21] that liquids adsorbed at a planar hard-wall exhibit the phenomenon of complete drying for all temperatures between the bulk triple and critical points. As the chemical potential $\mu$ of the liquid is reduced towards $\mu_{co}$, the value at liquid-gas coexistence, a thick film of gas intrudes between the liquid and the hard wall; the thickness of this film diverges in the limit $\mu \to \mu_{co}$. We re-visit this phenomenon within the framework of LMF, using an accurate hard-sphere DFT to treat the reference system, for the case of a hard-core fluid with a Yukawa attractive tail. In particular we examine the effective reference potentials that are required to describe the approach to complete drying. These potentials are repulsive but we find that the tails can decay in a monotonic or oscillatory fashion, depending on the temperature at which coexistence is reached.

The efficacy of the LMF theory depends on how intelligently the reference system is chosen. We examine the sensitivity of results of LMF to the division of the pair potential for a one-dimensional model that arises in the theory of effective interactions in colloid-polymer mixtures [22]. The effective colloid-colloid pair potential has a hard-core and a linear (attractive) tail of finite range. We choose the reference fluid to include the hard-core plus varying amounts of the tail. The remaining part of the tail is treated in mean-field fashion. Since for this model we can calculate the thermodynamics and structure (pair correlation function) of the uniform fluid exactly for any choice of reference system we ascertain how treating some or all of the tail within mean-field compares with exact results and provide an assessment of the limitations of the LMF approach.

Our paper is arranged as follows: In Sec.II we present our derivation of the LMF equation and provide a DFT perspective on the approximations that underpin this approach. Sec.III describes the study of complete drying at a planar hard wall while Sec.IV comments on strategies for dividing the pair potential and describes our calculations and results for the one-dimensional model. We conclude in Sec.V with further discussion of the relationship between LMF and DFT, pointing to the strengths and weaknesses of each approach.

## II. DERIVATION OF THE LMF EQUATION FROM DFT

Following [14] we begin with the exact YBG equation for the one-body density in the full system with pair potential $w(r)$:

$$k_B T \nabla \ln \rho^{(1)}(\mathbf{r};[\phi]) = -\nabla \phi(\mathbf{r}) - \int d\mathbf{r}' \rho(\mathbf{r}'|\mathbf{r}) \nabla w(|\mathbf{r}-\mathbf{r}'|) \qquad (3)$$

where $\phi(\mathbf{r})$ is the external potential and the conditional singlet density is given by

$$\rho(\mathbf{r}'|\mathbf{r}) \equiv \rho^{(2)}(\mathbf{r},\mathbf{r}')/\rho^{(1)}(\mathbf{r};[\phi]). \qquad (4)$$

Here $\rho^{(2)}(\mathbf{r},\mathbf{r}')$ is the usual pair distribution function [17]. Since Eq.(3) expresses force balance in the fluid it can be re-expressed in terms of the one-body direct correlation function $c^{(1)}(\mathbf{r})$. Recall that $c^{(1)}(\mathbf{r})$ satisfies

$$\Lambda^3 \rho^{(1)}(\mathbf{r};[\phi]) = \exp[\beta(\mu - \phi(\mathbf{r})) + c^{(1)}(\mathbf{r};[w])]. \qquad (5)$$

$\Lambda$ is the thermal de Broglie wavelength and $\beta = (k_B T)^{-1}$. $c^{(1)}$ is a functional of $\rho^{(1)}$, the one-body density, and we make explicit that it is also a functional of the pair potential. $-k_B T c^{(1)}$ is an effective one-body potential that combines with the external potential to determine self-consistently the equilibrium density profile of the fluid [4,17]. Taking the logarithm of (5) and differentiating one obtains the force balance equation [23]

$$k_B T \nabla \ln \rho^{(1)}(\mathbf{r};[\phi]) = -\nabla \phi(\mathbf{r}) + k_B T \nabla c^{(1)}(\mathbf{r};[w]) \quad (6)$$

which is completely equivalent to (3). In the absence of interactions $w=0$, $c^{(1)}=0$ and (6) reduces to the equation of hydrostatics for a non-interacting fluid in an external potential.

Consider now the reference or mimic fluid with pair potential $u_0(\mathbf{r})$ and external potential $\phi_R(\mathbf{r})$. It follows that the one-body density must satisfy the reference equivalent of (5):

$$\Lambda^3 \rho_R^{(1)}(\mathbf{r};[\phi_R]) = \exp[\beta(\mu_R - \phi_R(\mathbf{r})) + c_R^{(1)}(\mathbf{r};[u_0])] \quad (7)$$

where $\mu_R$ is the chemical potential of the reference fluid and $c_R^{(1)}$ is a functional of the one-body density and of $u_0(\mathbf{r})$. The analogue of (6) is

$$k_B T \nabla \ln \rho_R^{(1)}(\mathbf{r};[\phi_R]) = -\nabla \phi_R(\mathbf{r}) + k_B T \nabla c_R^{(1)}(\mathbf{r};[u_0]) \quad (8)$$

which expresses force balance in the reference fluid. We now invoke the equality (1), assuming that an effective potential $\phi_R(\mathbf{r})$ exists, and subtract (6) and (8). The logarithmic (ideal gas) terms cancel and it follows that

$$-\nabla \phi_R(\mathbf{r}) = -\nabla \phi(\mathbf{r}) + k_B T \nabla (c^{(1)}(\mathbf{r};[w]) - c_R^{(1)}(\mathbf{r};[u_0])). \quad (9)$$

This equation is exact provided $\phi_R(\mathbf{r})$ exists. It is completely equivalent to Eq.(7) in Ref. [14]. Note however that (9) remains valid for general interatomic potential functions. Since we have invoked only total force balance in the two systems (having identical density profiles), in the present treatment there is no need, at this stage, to assume a pair potential description for the potential functions in the full and reference systems.

One can easily integrate (9) to obtain

$$\phi_R(\mathbf{r}) = \phi(\mathbf{r}) - k_B T (c^{(1)}(\mathbf{r};[w]) - c_R^{(1)}(\mathbf{r};[u_0])) + C \quad (10)$$

where $C$ is a constant. Suppose that for vanishing external fields we have a uniform fluid of (bulk) density $\rho_b$. It then follows from the definition[17,23] of $c^{(1)}$ that

$$C = -(\mu_{ex}(\rho_b;[w]) - \mu_{exR}(\rho_b;[u_0])) \ . \tag{11}$$

Thus $C$ is the difference in excess chemical potential between the reference and the actual fluid at the given temperature.

As it stands Eq. (10) is formally exact but does not provide a means of formulating useful approximations. This changes if one adopts a DFT perspective. The one-body direct correlation function is given by [17, 23]

$$c^{(1)}(\mathbf{r};[w]) \equiv -\beta \frac{\delta F_{ex}[\rho;[w]]}{\delta \rho(\mathbf{r})} \tag{12}$$

where $F_{ex}[\rho;[w]]$ is the excess (over ideal) intrinsic Helmholtz free energy functional of the fluid described by the pair potential $w(\mathbf{r})$. $F_{ex}[\rho;[w]]$ is a unique functional of the one-body density $\rho(\mathbf{r}) \equiv \rho^{(1)}(\mathbf{r};[w])$; it is the same functional for all external fields [17,23]. Similarly one can introduce the excess intrinsic Helmholtz free energy functional of the reference fluid $F_{exR}[\rho;[u_0]]$ and write Eq. (10) as

$$\phi_R(\mathbf{r}) = \phi(\mathbf{r}) + \frac{\delta}{\delta \rho(\mathbf{r})}(F_{ex}[\rho;[w]] - F_{exR}[\rho;[u_0]]) + C, \tag{13}$$

which remains formally exact.

The standard mean-field treatment of attractive forces within DFT [3, 4] approximates the difference in excess free energy functionals as

$$F_{ex}[\rho;[w]] - F_{exR}[\rho;[u_0]] = \frac{1}{2}\iint d\mathbf{r} d\mathbf{r}' \rho(\mathbf{r})\rho(\mathbf{r}')u_1(|\mathbf{r}-\mathbf{r}'|) \ . \tag{14}$$

In this approximation (13) reduces to

$$\phi_R(\mathbf{r}) = \phi(\mathbf{r}) + \int d\mathbf{r}' \rho(\mathbf{r}')u_1(|\mathbf{r}-\mathbf{r}'|) + C, \tag{15}$$

where the boundary condition will determine $C$, the constant of integration. For a fluid adsorbed at a wall one can require the external potentials to vanish far from the wall and the density profile to approach the bulk value: $\rho(\mathbf{r}) \to \rho_b$. Then $C = -\rho_b \int d\mathbf{r}' u_1(r')$ which is precisely the same result that follows from (11) using (14). Eq.(15) is identical to the LMF equation- see Eq.(8) in Rodgers and Weeks[14] and Eq. (4) in Weeks et.al. [6], where $\phi_R(\mathbf{r})$ is termed the effective reference field (ERF). It is important to note that in Eq.(15) $\rho(\mathbf{r}') \equiv \rho_R(\mathbf{r}';[\phi_R])$ and, as emphasized by Weeks and co-workers, in the language of distribution functions one has succeeded in truncating the hierarchy since $\phi_R(\mathbf{r})$ depends only on the one-body density $\rho_R(\mathbf{r};[\phi_R])$. In implementing LMF one must solve (15) self-consistently for the reference (or mimic) system described by a (short ranged) pair potential $u_0(r)$. Most, but not all [6], implementations employ computer simulations.

Weeks and co-workers [6, 14] arrive at the approximation (15) via very insightful arguments about the nature of the conditional singlet densities $\rho(\mathbf{r}'|\mathbf{r})$ in the full and mimic systems; specifically they expect these to be similar at short range. Within the present DFT approach the LMF equation follows directly from (14). This constitutes the simplest possible treatment of attractive forces: no correlations are included as an inhomogeneous pair correlation function is replaced by a product of one-body densities [4,23]; see Eqs. (42, 43) in Sec.V. Another way of understanding the physical content of (14) is to take two functional derivatives w.r.t. density to obtain the pair direct correlation functions [4,17,23]. One finds

$$c^{(2)}(\mathbf{r},\mathbf{r}';[w]) = c^{(2)}_{RPA}(\mathbf{r},\mathbf{r}';[w]) \equiv c^{(2)}_R(\mathbf{r},\mathbf{r}';[u_0]) - \beta u_1(|\mathbf{r}-\mathbf{r}'|). \tag{16}$$

For a uniform fluid of constant density $\rho_b$ we recognize immediately that (16) is simply a random phase approximation (RPA) for the attractive or, more generally, for the tail contribution to the pair direct correlation function.

Suppose now that we adopt the DFT perspective and we are able to choose a reference system R where $F_{exR}[\rho;[u_0]]$ is known exactly. The grand potential functional corresponding to (13) is

$$\Omega[\rho] = F_{id}[\rho] + F_{exR}[\rho;[u_0]] + \frac{1}{2}\iint d\mathbf{r} d\mathbf{r}' \rho(\mathbf{r})\rho(\mathbf{r}')u_1(|\mathbf{r}-\mathbf{r}'|) + \int d\mathbf{r}\rho(\mathbf{r})(\phi(\mathbf{r}) - \mu) \tag{17}$$

where $F_{id}[\rho]$ is the ideal gas contribution. On minimization (17) yields the Euler-Lagrange equation

$$k_B T \ln(\Lambda^3 \rho(\mathbf{r})) + \frac{\delta}{\delta\rho(\mathbf{r})} F_{exR}[\rho;[u_0]] + \int d\mathbf{r}' \rho(\mathbf{r}')u_1(|\mathbf{r}-\mathbf{r}'|) + \phi(\mathbf{r}) = \mu \tag{18}$$

which can be re-expressed, using the LMF equation (15), as

$$k_B T \ln(\Lambda^3 \rho(\mathbf{r})) + \frac{\delta}{\delta\rho(\mathbf{r})} F_{exR}[\rho;[u_0]] + \phi_R(\mathbf{r}) = \mu_R \tag{19}$$

where $\mu_R$ is the chemical potential of the reference fluid and we have used the result that the chemical potential of the fluid with the full potential and of density $\rho_b$ is

$$\mu = \mu_{id}(\rho_b) + \mu_{exR}(\rho_b) + \rho_b \int d\mathbf{r}' u_1(r') \quad (20)$$
$$\equiv \mu_R - C.$$

Eq. (19) is simply the Euler–Lagrange equation for the *reference* fluid subject to an external potential $\phi_R(\mathbf{r})$; it is identical to (7). We could make this even more explicit by making the substitution $\rho(\mathbf{r}) \equiv \rho_R(\mathbf{r};[\phi_R])$. Solving (19) with the exact functional $F_{exR}[\rho;[u_0]]$ is completely equivalent to solving (15) using simulations.

Of course one does not have an exact functional for a three dimensional fluid and in practice within DFT one resorts to using an approximate intrinsic free energy functional for the reference system. Very accurate DFTs have been developed for hard-sphere models [1-3] so a natural choice of reference system sets $u_0(r)$ equal to the hard-sphere potential. For a pair potential $w(r)$ with a hard core and an (attractive) tail $u_1(r)$ one can then solve (19) using, for example, a Fundamental Measures Theory for the hard-sphere functional $F_{ex}^{HS}[\rho]$. This procedure should be almost as accurate as using Monte Carlo simulation to solve (15) for the hard-sphere fluid. However, we emphasize that this LMF based strategy is completely equivalent to DFT with the standard mean-field treatment (14) of the tail, i.e. results for the density profile and thermodynamics will be identical to those obtained by solving the usual Euler-Lagrange equation (18) with $F_{exR}[\rho;[u_0]] = F_{ex}^{HS}[\rho]$. In Sec.III we shall implement this particular strategy for the case of a hard-core Yukawa liquid adsorbed at a hard-wall and investigate in detail the form of the effective reference potential required to describe drying.

## III. EFFECTIVE REFERENCE POTENTIAL FOR DRYING AT A HARD-WALL.

As we remarked in the Introduction, providing a reliable description of complete drying requires a quantitative theory of both repulsive and attractive contributions to the free energy. An acceptable theory should account for the following : i) if the bulk fluid far from the substrate is in a dense liquid state well-removed from bulk gas-liquid coexistence the density profile must exhibit oscillations with a period roughly equal to the atomic diameter and ii) for liquid states very close to bulk coexistence the attractive interatomic forces must ensure that a 'gas-like' film with low density develops close to the wall so that the density profile becomes a composite of the wall-gas and gas-liquid interfaces [18-21, 24]. More precisely, for a fluid with short-ranged interatomic potentials of finite support or with exponential or faster decaying tails, the thickness of the drying film diverges as $\ln(\delta\mu)$ and the wall-liquid surface tension acquires a non-analytic contribution, i.e. $\gamma_{wl} = \gamma_{wg} + \gamma_{gl} + A\delta\mu\ln(\delta\mu)$ where $\gamma_{wg}$ is the wall-gas tension, $\gamma_{gl}$ is the gas-liquid tension, $A$ is a constant and $\delta\mu \equiv \mu - \mu_{co}$ is the deviation from coexistence [21, 24].

Here we examine complete drying at a hard-wall in the context of LMF, enquiring how the effective reference potential $\phi_R(\mathbf{r})$ varies on the approach to coexistence. We consider a model fluid described by a hard-core Yukawa potential:

$$\begin{aligned} w_Y(r) &= \infty & r \leq \sigma \\ &= -\varepsilon\frac{\sigma}{r}\exp[-Z(r/\sigma - 1)] & r > \sigma \end{aligned} \quad (21)$$

where $\sigma$ is the hard-sphere diameter, $\varepsilon$ is the well-depth and $Z$ is a (dimensionless) inverse range parameter. For the reference system we take the hard-sphere fluid of diameter $\sigma$ and treat this within Rosenfeld's [25] Fundamental Measures Theory (FMT). FMT is known to describe accurately the density profile and surface tension of hard-sphere fluids adsorbed at hard-walls for bulk densities approaching those at bulk freezing $\rho_f\sigma^3 = 0.94$. More sophisticated versions of FMT fare slightly better at very high packing fractions [1-4].

In our DFT calculations we minimize (17) with the reference excess free energy functional now given by the Rosenfeld hard-sphere functional and the external potential is simply that of the hard-wall:

$$\begin{aligned} \phi(\mathbf{r}) \equiv \phi_{hw}(z) &= \infty & z \leq \sigma/2 \\ &= 0 & z > \sigma/2. \end{aligned} \quad (22)$$

We choose the attractive part of the pair potential to be

$$\begin{aligned} u_{1Y}(r) &= -\varepsilon & r \leq \sigma \\ &= w_Y(r) & r > \sigma. \end{aligned} \quad (23)$$

It is well-known that when a hard-core model is treated within the RPA the attractive potential is not defined uniquely in the core. Our choice (23) yields a bulk gas-liquid coexistence curve (binodal) within RPA that is in reasonable agreement with the results from the more accurate Mean-Spherical Approximation (MSA) [26]. Using Picard iteration we solve the (one-dimensional) Euler-Lagrange equation (18) for the equilibrium density profile $\rho(z) \equiv \rho(\mathbf{r})$ at different state points approaching the binodal. We note that the functional described by (17) yields equilibrium profiles that satisfy the Gibbs adsorption equation and the hard-wall contact theorem $\rho(\sigma/2) = \beta p$, where $p$ is the pressure of the bulk fluid [4]. The effective reference potential, which is also one-dimensional: $\phi_R(\mathbf{r}) \equiv V_{eff}(z)$, is obtained by inserting the equilibrium density profile into the LMF equation (15). We repeat that this procedure is precisely the same as solving (19) self-consistently for $V_{eff}(z)$.

The first case that we consider has $Z=1$: the Yukawa tail is fairly long-ranged. In Fig.1 we show the gas-liquid binodal and the corresponding spinodal calculated for a bulk fluid described by the functional (17), i.e. by the RPA treatment of the attractive tail. A comparison with MSA results for $Z=1$ is given in Fig.2 of Ref. [26]. We choose two paths to the binodal. The first fixes the bulk density (far from the wall) at $\rho\sigma^3=0.65$ and reduces the temperature. Results for density profiles are shown in Fig. 2(a). For $\beta\varepsilon=0$, the system is a hard-sphere fluid at a hard-wall and the profile exhibits oscillations, albeit with exponentially decreasing amplitude, as $z \to \infty$. For this system at this bulk density FMT is extremely accurate [1]. We expect a similar degree of accuracy for the same hard-sphere fluid subject to effective potentials. As $\beta\varepsilon$ is increased the contact value $\rho(\sigma/2)$ is reduced. For $\beta\varepsilon= 0.4$ the profile still exhibits oscillations at short distances but for the larger values of $\beta\varepsilon$ the oscillations are eroded. For $\beta\varepsilon = 0.5290$, close to the value at the binodal $\beta\varepsilon=0.5293$, the density profile (dash-dot line) exhibits an intruding gas-film and a fairly broad gas-liquid interface. It is clear that we have the usual complete drying scenario.

Fig. 2(b) displays the corresponding effective potentials $V_{eff}(z)$. For $\beta\varepsilon=0$ (hard-spheres) the effective potential is simply the hard-wall potential (22). As $\beta\varepsilon$ increases $V_{eff}(z)$ develops a repulsive tail that becomes larger and more extended as the binodal is approached. At $\beta\varepsilon=0.5290$ $V_{eff}(\sigma/2)$ has a value of about 10 $k_BT$ and it decays with a range set by the form of the density profile. Such a shape for $V_{eff}(z)$ is consistent with the intuitive picture of the hard-sphere liquid being 'pushed away' from the wall by an additional effective repulsive potential creating the dry region. We note that apart from $\beta\varepsilon=0$ all the density profiles increase monotonically to the bulk liquid value. Similarly the effective potentials decay monotonically (exponentially) to zero.

In Fig.3 we show results for a fixed bulk density $\rho\sigma^3=1.0$. The sequence of density profiles is similar to those at the lower density in that a drying film of increasing thickness develops as $\beta\varepsilon$ approaches the binodal. However the gas-liquid interface is sharper since the temperature is now far below that of the critical point. Note that the profiles are weakly oscillatory even for $\beta\varepsilon=1.134348$ (dash-dot line), a value very close to that at the binodal $\beta\varepsilon=1.134349$. The effective potentials (Fig. 3(b)) exhibit a similar trend to those in Fig.2(b) but with values of $V_{eff}(\sigma/2)$ as large as 32 $k_BT$. Closer inspection, on the expanded scale of the inset, reveals that the asymptotic decay into bulk (large $z$) is non-trivial. For $\beta\varepsilon=1.0$ one clearly observes (solid curve) monotonic exponential decay up to about $13\sigma$ followed by (exponentially damped) oscillatory decay at larger distances. For larger values of $\beta\varepsilon$, approaching the binodal, the effective potentials appear to decay in a monotonic exponential fashion until $z/\sigma\sim 20$ when noise sets into the numerical results making it difficult to discern the asymptotic behaviour. These results might appear counter-intuitive until one recalls the factors that determine the asymptotic decay (into bulk) of wall-fluid density profiles [27]. We postpone discussion until later after we present the results for the second choice of Yukawa tail, namely $Z=3$. This corresponds to a shorter ranged attractive interaction than the first model.

The RPA phase diagram is shown in Fig.4. As expected the critical temperature is considerably lower for this value of $Z$. Within the RPA the critical density $\rho_c\sigma^3=0.2457$ is the same for all $Z$ [26]. Once more we choose two paths at constant bulk density. Fig.5 is for the path with $\rho\sigma^3=0.5$ which intersects the binodal at $\beta\varepsilon=1.31279$. The density profiles in Fig.5(a) are similar to those in Fig. 2(a). For the hard-sphere fluid, $\beta\varepsilon=0$, the oscillations are less pronounced and the contact value $\rho(\sigma/2)$ is smaller since the bulk density is smaller than in Fig.2 (a).The state with $\beta\varepsilon=1.0$ is supercritical but the density profile exhibits weak oscillations at short distances. For values of $\beta\varepsilon$ approaching the binodal the profiles increase monotonically with distance $z$. The corresponding effective potentials are shown in Fig.5(b) where the inset shows that $V_{eff}(z)$ decays exponentially into bulk. For the three state points closest to the binodal the exponential decay lengths extracted from the plots are almost identical. In Fig.6(a) we show the results for the profiles on the path with a larger bulk density $\rho\sigma^3=0.9$ which intersects the binodal at $\beta\varepsilon=2.604651$. For $\beta\varepsilon=2.0$ the density profile (solid line) exhibits pronounced oscillations whose amplitude decays quite slowly into bulk. Oscillations also persist on the liquid side of the gas-liquid interface as $\beta\varepsilon$ increases towards its value at the binodal and the thickness of the gas film increases. The effective potentials corresponding to these profiles are displayed in Fig.6(b). For $\beta\varepsilon=2.0$ (solid line) there is weak oscillatory structure in the range 1 to $3\sigma$ and at larger distances (see inset) $V_{eff}(z)$ exhibits exponentially damped oscillations with a period of about 1 $\sigma$. At larger values of $\beta\varepsilon$, approaching the binodal, $V_{eff}(z)$ decays monotonically until $z$ is in the tail region and then exponentially damped oscillations are clearly visible. These oscillations set in at distances that increase with $\beta\varepsilon$, i.e. their onset is governed by the thickness of the film of gas. For the four values of $\beta\varepsilon$ closest to the binodal the exponential decay lengths obtained from fitting the envelope are the same. As the decay length should be an intrinsic property of the bulk liquid and these bulk state points are very close this observation gives us confidence that our numerics are robust for distances up to about 20 $\sigma$. It is clear that the tails of the effective potentials for $\rho\sigma^3=0.9$ have a very different character from those for $\rho\sigma^3=0.5$.

What is the origin of such behaviour? The general theory [27] of the asymptotic decay of one-body densities argues that for fluids with short-ranged interatomic pair potentials the wall-fluid density profile $\rho(z)$ should approach its bulk value $\rho_b$ as $z \to \infty$ in the same way that $r(g(r)-1)$ approaches zero as $r \to \infty$ *provided* the wall-fluid external potential is of finite range. $g(r)$ is the radial distribution function of the bulk fluid of density $\rho_b$. This suggests that in interpreting our results we should consider the location of the Fisher-Widom (FW) line in the bulk phase diagram [27]. The FW line marks cross-over from one type of asymptotic decay to another. Specifically on the high density side of the FW line the ultimate decay of $r(g(r)-1)$ is exponentially damped oscillatory whereas on the low density side it is monotonic (exponential). For state points close to the FW line

$$r(g(r)-1) \sim B\exp(-\alpha_0 r) + B_{osc}\exp(-\alpha_0^{osc} r)\cos(\alpha_1 r + \theta) \quad , r \to \infty \qquad (24)$$

where $i\alpha_0$ is the pure imaginary and $\alpha = \alpha_1 + i\alpha_0^{osc}$ is a complex pole of the structure factor with the smallest imaginary parts [27]. $\theta$ is a phase angle and $B$ and $B_{osc}$ are amplitudes. If $\alpha_0 < \alpha_0^{osc}$ the ultimate decay is monotonic whereas if $\alpha_0 > \alpha_0^{osc}$ it is damped oscillatory. The FW line is defined as the locus of points where $\alpha_0 = \alpha_0^{osc}$. Thus we expect that in the present model the asymptotic behaviour of the wall-fluid density profile should be accounted for by the two-pole approximation

$$\rho(z) - \rho_b \sim D\exp(-\alpha_0 z) + D_{osc}\exp(-\alpha_0^{osc} z)\cos(\alpha_1 z + \tilde{\theta}), \quad z \to \infty \qquad (25)$$

in an obvious notation. FW lines have been calculated for a variety of model fluids, including the square-well [27] and truncated LJ fluid [28]. Generally the FW lines intersect the binodal on the liquid side and are bounded at high density by the (liquid) spinodal. In the limit $T \to \infty$ the FW line approaches $\rho = 0$ asymptotically reflecting the fact that the hard sphere $g(r)$ exhibits oscillations for all non-zero densities [28]. The precise location of the FW line in the $\rho$-$T$ diagram depends on the particular model fluid.

Brown et.al. [26] calculated the poles of the structure factor and determined the FW line for the present hard-core Yukawa model using both the RPA and the MSA. Although these authors do not provide explicit results for $Z=3$ it is expected from their analysis that within RPA the FW line intersects the liquid binodal at a density lying between $\rho\sigma^3=0.9$ and 0.5. We can safely assume that the results presented in Fig.5 for $\rho\sigma^3=0.5$ and $\beta\varepsilon > 1$ correspond to states on the low density (monotonic) side of the FW line and that the first term in (25) dominates. Given that the density profile should approach the bulk value monotonically it is not surprising that the effective potential is also monotonically decaying; the latter is a convolution of the profile $\rho(z')$ with the x-y integral of the attractive potential $u_1$-see Eq. (15). On the other hand, for $\rho\sigma^3=0.9$ we expect all the states considered in Fig.6 to lie on the oscillatory side of the FW line and the second term in (25) to dominate. The convolution yields exponentially damped oscillatory effective potentials for large distances. We return now to the results for $Z=1$. For this case the FW line is plotted in the $\rho$-$T$ plane in Fig.2 of Ref. [26]. The RPA and MSA results are very close. The FW line intersects the RPA liquid binodal at $\rho\sigma^3 \sim 0.86$ and $k_B T/\varepsilon \sim 1.26$. This means that the although the state points considered here in Fig.3 should lie on the oscillatory side these are not far removed from the FW line -recall the latter is bounded at high densities by the liquid spinodal. One might reasonably expect both term in (25) to contribute. Moreover there is no theory for the amplitudes. We surmise that this is the reason for the behaviour observed in Fig. 3(b). The ultimate decay of $V_{eff}(z)$ might well be damped oscillatory for these state points.

One might argue (justifiably) that some of the bulk densities we consider are unphysically large. The Rosenfeld FMT we use does not account for freezing. For example, $\rho\sigma^3=1.0$ is greater than the freezing density of the hard-sphere fluid and the triple point density of the hard-core Yukawa model with $Z=1.0$. Monte Carlo simulations[29] show that the same model with $Z=3.9$ exhibits gas-liquid coexistence in only a narrow temperature range and that the triple point density $\rho_t\sigma^3 \sim 0.75$. Thus for $Z=3$ a density of $\rho\sigma^3=0.9$ is unlikely to be in the liquid phase. Nevertheless the features that we find in the tail of $V_{eff}(z)$ should be found in equivalent treatments of other models where the FW line intersects the binodal closer to the critical point and therefore at much lower densities; examples are the square –well fluid[27] and a hard-core with a truncated LJ tail [28].

Our results demonstrate that for complete drying at a hard-wall the LMF effective reference potential $\phi_R(\mathbf{r}) \equiv V_{eff}(z)$ must exhibit several important features if the hard-sphere reference system is to capture the details of the density profiles that characterize the drying phenomenon occurring in the fluid with the full pair potential. Some of the subtlety can be appreciated by recalling that for any hard-sphere fluid subject to a wall-fluid potential of finite range the density profiles must decay to bulk in exponentially damped oscillatory fashion. There is no pure imaginary pole so the first term on the r.h.s. of (25) is absent. The second term corresponds to the oscillatory pole of the structure factor of the bulk hard-sphere fluid having the smallest imaginary part. In order to account for drying a $V_{eff}(z)$ must be constructed (self-consistently) so that the hard-sphere fluid feels i) the correct repulsive potential hump to produce the appropriate film of gas and ii) the correct tail to produce a density profile that contains both terms of (25) with the inverse decay lengths $\alpha_0$ and $\alpha_0^{osc}$ and the wavelength $2\pi/\alpha_1$ now corresponding to the bulk liquid with the full pair potential $w(r)$. We have demonstrated that within the DFT implementation it is possible to achieve such a $V_{eff}(z)$. But we achieved this by solving (18) iteratively using the explicit Rosenfeld FMT functional for hard-spheres and then substituting the resulting equilibrium density profile into (15) to obtain $V_{eff}(z)$. As we noted earlier this is equivalent to solving (19) iteratively.

Suppose now we switch from the DFT description of the hard-sphere reference system and decide to use computer simulation to solve the LMF equation (15) self-consistently for the same hard-sphere system. We should expect to find the same results; FMT is very accurate. However, obtaining the level of detail we found in DFT constitutes a challenge. With considerable computational effort one might succeed in determining the correct form of the potential hump but to ascertain the correct details of the tail of $V_{eff}(z)$ at state points close to the binodal is certainly a tall order for simulation.

We conclude this Section by emphasizing that the present mean-field DFT treatment of attractive interatomic forces (Eqs. 14 & 17) does not include effects of capillary wave-fluctuations. This was recognized and discussed at some length in an earlier paper on drying [27]. We re-iterate the failings. As $\delta\mu \to 0$, the thickness of the film of gas increases, the gas-liquid interface moves further from the hard-wall and the capillary-wave fluctuations become more pronounced. There are three main effects: i) The fluctuations produce a broadening of the gas-liquid interface that is not captured by the DFT. ii) The fluctuations renormalize the thickness $l$ of the drying (gas) film which should diverge as

$$l = \xi_b(1+\omega/2)\ln(a\delta\mu), \qquad \delta\mu \to 0 \qquad (26)$$

where $\omega = k_B T/(4\pi\gamma_{gl}\xi_b^2)$ is the usual dimensionless parameter characterizing the strength of interfacial fluctuations. Mean-field (DFT) theory corresponds to a very stiff interface with $\omega = 0$. $a$ is a constant and $\xi_b$ is the true correlation length in the wetting phase. In the present case of drying $\xi_b$ is the correlation length in the bulk gas at chemical potential $\mu_{co}$. iii) The fluctuations erode any oscillations arising in the mean-field density profile such as those calculated in DFT. Incorporating the effects of capillary wave fluctuations is non-trivial and for the chemical potential deviations $\delta\mu$ considered here one can only speculate that the film thickness and the width of the gas-liquid interface are not altered grossly. However, the degree of erosion of the oscillations in the profile on the liquid side of the interface is very sensitive to the value of $\omega$ [27, 30].

It should be clear that a LMF treatment of drying, using say simulation for the reference fluid, will suffer *precisely* the same shortcomings as our DFT regarding the incorporation of the effects of capillary wave fluctuations. The shortcomings of the LMF approach for the free gas-liquid interface are clearly expounded by Katsov and Weeks [31] and by Weeks [9]. We return to the problem of drying on a hard-sphere solute, treated within LMF by Katsov and Weeks [8], in the concluding Section.

**IV. CHOICE OF REFERENCE FLUID: DIVISION OF THE POTENTIAL *w*(*r*)**

The strategy behind LMF is that of dividing the full pair potential into a reference part, which can be treated by simulation without extreme computational cost or by some alternative very accurate procedure, and a remainder which is treated within the mean-field approximation. As mentioned in the Introduction, for both uniform and non-uniform liquids the usefulness of LMF depends on making an appropriate choice of the reference fluid. The choice is usually dictated by the physics of the particular system. For example, in the case of ionic liquids Weeks and co-workers [10,14] have made remarkable progress by splitting the Coulomb potential as $1/r = u_0(r) + u_1(r)$, with $u_0(r) = erfc(r/\Gamma)/r$ and $u_1(r) = erf(r/\Gamma)/r$, where $\Gamma$ is a smoothing length. $u_1(r)$ is slowly varying over the range $\Gamma$, which includes the long range tail of the potential but is finite for all values of *r*. $u_0(r)$ is shorter ranged and describes the strong repulsive part of the Coulomb potential for small *r*. The parameter $\Gamma$ defines the distance at which $u_0(r)$ is smoothly truncated. Choosing $\Gamma$ too small is likely to produce poor results in the LMF treatment whereas choosing this parameter too large is equivalent to requiring a long-ranged reference fluid which defeats the purpose of making the division. Rodgers and Weeks [14] describe how to make an appropriate choice for $\Gamma$.

Making an effective division of the pair-potential has, of course, a long history in liquid state physics especially in the context of perturbation theories for uniform fluids [17]. However, what is interesting and useful about this way of splitting the potential is that the potential has not been cut at a certain distance from the core, but that there is a *smooth* crossover as *r* is increased, with an increasing proportion of the tail of the potential entering $u_1(r)$. This leads us to believe that there should be other ways to split the potential, not considered before, that would in the context of the LMF approach allow us to tackle (inhomogeneous) liquid state problems more efficiently.

In the following we consider two aspects of the division (2) of the full pair potential into a reference part $u_0(r)$ and a remainder $u_1(r)$. Although we focus on properties of the uniform fluid we emphasize that the LMF theory can be applied to both uniform and non-uniform situations.

### A. A Lennard-Jones type of potential

The challenge is to find new and intelligent ways to make the division in Eq. (2). For example, for a fluid of particles interacting via the LJ pair potential $w_{LJ}(r) = 4\varepsilon((\sigma/r)^{12} - (\sigma/r)^6)$, one might define the division as follows:

$$u_0(r) = 4\varepsilon((\sigma/r)^{12} - (1-\delta)(\sigma/r)^6) \text{ and } u_1(r) = -4\varepsilon\delta(\sigma/r)^6, \qquad (27)$$

with $0 \leq \delta \leq 1$. Choosing $\delta=1$ is an obvious splitting, but choosing other values may also be interesting and useful. For example, consider the case of a uniform fluid, with the full potential $w_{LJ}(r)$, that is at a state point close to where a phase transition occurs. It is at such state points where integral equation theories can often break down or be very difficult to solve. Choosing $\delta$ small can lead to the reference fluid being a little away from the phase transition (say supercritical). However, because $\delta$ is small, the change in the fluid structure in going from the reference system to the full system should not be very great and thus LMF theory should be rather accurate-albeit still performing at mean-field level.

Another idea worthy of pursuing is to not to split the given pair potential but instead to introduce a fictitious smooth and slowly varying potential $v(r)$ that one *adds* to the true system potential $w(r)$ forming a reference system whose particles interact via the potential $u_0(r) = w(r) + v(r)$. The potential $v(r)$ should be chosen so that the structure and thermodynamics of this reference system is more amenable to theoretical treatments or simulation than the true system. The effect of the fictitious additional contribution $u_1(r) = -v(r)$ is then to be removed via the LMF approach.

### B. An exactly solvable one-dimensional model

One-dimensional models have an important role in statistical physics as it is often possible to obtain exact solutions for the thermodynamic properties and for correlation functions in such models enabling one to examine the accuracy and hence the reliability of approximation schemes along the lines of those proposed above. We adopt this strategy here and consider a one-dimensional model described by the pair-potential

$$\beta w(x) = \infty \qquad |x| \leq \sigma$$
$$= \beta w_{tail}(x) \qquad |x| > \sigma \qquad (28)$$

where

$$\beta w_{tail}(x) = 0 \qquad\qquad |x| \leq \sigma$$
$$= -z_p(\sigma + \sigma_p - |x|) \qquad \sigma < |x| \leq \sigma + \sigma_p \qquad (29)$$
$$= 0 \qquad\qquad |x| > \sigma + \sigma_p.$$

|x| is the distance between the centres of colloidal rods. This potential arises in the context of colloid-polymer mixtures; it is the one-dimensional analogue of the celebrated Asakura-Oosawa result for the effective colloid-colloid potential of hard-sphere colloids immersed in ideal polymer. Eq.(29) is derived [22] for a mixture of hard-rods of length $\sigma$ and (ideal)polymer coils that are mutually non-interacting. The centre of each polymer coil is excluded from the centre of the colloid by a distance $(\sigma + \sigma_p)/2$ and one can regard $\sigma_p$ as the 'length' of the polymer coil. By formally integrating out the degrees of freedom of the polymer in a semi-grand canonical ensemble one obtains (29). $z_p$ is the fugacity of the polymer and since this is ideal $z_p = \rho_p^r$, the number density of polymer in the reservoir. It was shown in Ref.[22] that provided the ratio $\sigma_p/\sigma < 1$ the effective Hamiltonian for the one-dimensional mixture has no three or higher-body terms. However, for our present purposes we may simply regard (28, 29) as defining a convenient model whose properties we investigate using different divisions of the pair potential.

We choose the following division depending on the parameter $\lambda$:

$$w(x) = u_0(\lambda; x) + u_1(\lambda; x) \qquad (30)$$

with reference potential

$$u_0(\lambda;x) = \infty \qquad x \leq \sigma$$
$$= (1-\lambda)w_{tail}(x) \qquad x > \sigma \qquad (31)$$

and remaining potential

$$u_1(\lambda;x) = \lambda w_{tail}(x) \qquad (32)$$

with $0 \leq \lambda \leq 1$. Setting $\lambda = 1$ treats all of the (attractive) tail as a remainder whereas when $\lambda = 0$ the reference system describes the full potential $w(x)$. Intermediate values of $\lambda$ apportion a certain piece of the tail to the reference potential and the remaining piece to $u_1(\lambda;x)$.

The advantage of the one-dimensional model is that for the case of a uniform fluid we can determine the properties of the reference fluid exactly for any value of $\lambda$ in the range $0 \leq \lambda \leq 1$. (Of course this is not specific to the present model; we could consider many other models with nearest-neighbour interactions.)

We focus first on the equation of state which is easily calculated by the standard Laplace transform method of Takahashi [32]. For the full potential the number density of colloidal rods $\rho$ is given as a function of pressure $P$ by the equation (see (A.10) of Ref.[22])

$$\rho = -\frac{\tilde{J}(\beta P)}{\tilde{J}'(\beta P)} \qquad (33)$$

where

$$\tilde{J}(s) = \int_0^\infty dx\, e^{-sx} e^{-\beta w(x)} \quad . \qquad (34)$$

The same formulae pertain to a reference fluid with pair potential $u_0(\lambda;x)$, and $0 \leq \lambda \leq 1$. We denote the corresponding pressure $P_{0\lambda}$.

Suppose now we choose a value of $\lambda > 0$ and specify $P_{0\lambda}$. The density $\rho$ is then determined. We now treat $u_1(\lambda;x)$, the remaining contribution to the potential, in mean-field approximation, i.e. using Eq.(14). It follows that the (mean field) pressure of the full (one-dimensional) system with potential $w(x)$ depends on $\lambda$ and is given by

$$P_{MF\lambda} = P_{0\lambda} + \frac{\rho^2}{2} \int_{-\infty}^{\infty} dx u_1(\lambda; x) \tag{35}$$

which reduces to

$$\begin{aligned} P_{MF\lambda} &= P_{0\lambda} + \rho^2 \lambda \int_{\sigma}^{\infty} dx w_{tail}(x) \\ &= P_{0\lambda} - k_B T \rho^2 \lambda z_p \sigma_p^2 / 2. \end{aligned} \tag{36}$$

The specified pressure of the reference system is reduced by an amount proportional to the integrated strength of the portion of the tail potential that is being treated in mean-field approximation.

In Fig.7 we show results for the reduced pressure $\beta P \sigma$ of the full system as a function of the reduced colloid density $\rho \sigma$ for three different values of the polymer fugacity $z_p$. Note that the full pair potential at contact is $w(\sigma) = -k_B T z_p \sigma_p$. Increasing $z_p$ increases the depth of the attractive well. The size ratio is fixed at $\sigma_p / \sigma = 0.9$, one of the ratios considered in Ref. [22]. For each value of $z_p$ we plot results for four values of the parameter $\lambda$ that specifies the division of the potential. The black solid curves in each figure are the exact results; these have $\lambda=0$. In Fig.7(a), $z_p \sigma =1$, treating 10% of the tail potential in mean-field (red dashed line with $\lambda=0.1$) is an excellent approximation. However, as $\lambda$ increases one finds sizeable deviations from the exact result. For $z_p \sigma =5$ the result for $\lambda=0.1$ is still a very good approximation to the exact but the deviations for larger $\lambda$ are very large, as shown in Fig. 7(b). For $z_p \sigma =9$, which corresponds to a very strongly attractive tail potential, Fig. 7(c) shows that treating 10% of the tail in mean-field ($\lambda=0.1$) already leads to an unphysical result. In this one-dimensional system with nearest-neighbour interactions the pressure should increase monotonically with density; there is no phase transition. We see that for $\lambda=0.1$ (red dashed line) the pressure is negative at some intermediate densities and there is a van der Waals loop. For $\lambda=0.5$ the pressure becomes more negative with the minimum shifting to lower density. Finally treating all the tail in mean field ($\lambda=1.0$), which is the standard RPA , shifts the minimum further and the pressure remains positive for all densities but very far removed from the exact result. Note that for this value of $z_p \sigma$ there is a critical value of $\lambda$, small and $< 0.1$, where $(\partial P / \partial \rho)_{z_p}$ first becomes zero.

It is important to recognize that the present exercise constitutes a very stringent test of any mean-field approximation. For the largest values of $z_p\sigma$ the potential well depths are many $k_BT$ and the inverse compressibility

$$\chi^{-1} \equiv \rho(\partial P/\partial \rho)_{z_p} \tag{37}$$

is very small for densities up to about $\rho\sigma = 0.8$. At higher densities the pressure increases rapidly and, for all $z_p\sigma$, diverges in the close packing limit $\rho\sigma \to 1$. Interestingly the mean-field treatment with $\lambda=0.1$ does appear to capture the correct steep increase-as does a different (free-volume) approximation [22]. However, the latter suffers the same defect as the present in the lower density regime.

We focus next on the pair correlation function of the uniform fluid. Specifically we calculate the static structure factor $S(k)$ using the same division (30-32) of the pair potential. The structure factor $S_{0\lambda}(k)$ can be calculated exactly for a reference potential $u_0(\lambda;x)$ using the result

$$S_{0\lambda}(k) = \frac{1-\exp[j(P_{0\lambda}+ik)-2j(P_{0\lambda})+j(P_{0\lambda}-ik)]}{(1-\exp[j(P_{0\lambda}+ik)-j(P_{0\lambda})])(1-\exp[j(P_{0\lambda}-ik)-j(P_{0\lambda})])} \tag{38}$$

where $j(s) = \ln \tilde{J}(s)$ and $P_{0\lambda}$ is the pressure of the reference fluid. This result pertains to any one-dimensional fluid with nearest-neighbour interactions (see (6.4) of Ref. [33]). In order to calculate the structure factor we first specify $P_{0\lambda}$. This determines the density via (33). We then use the Ornstein-Zernike relation to obtain $\hat{c}_{0\lambda}^{(2)}(k)$, the Fourier transform of the pair direct correlation function of the reference fluid:

$$1-\rho\hat{c}_{0\lambda}^{(2)}(k) = [S_{0\lambda}(k)]^{-1}. \tag{39}$$

The pair direct correlation function in the mean-field treatment follows from the same treatment that lead to (16). This function depends on $\lambda$ and is given by:

$$\begin{aligned}\hat{c}_{MF\lambda}^{(2)}(k) - \hat{c}_{0\lambda}^{(2)}(k) &= -\beta\hat{u}_1(\lambda;k) \\ &= -\beta\lambda\hat{w}_{tail}(k) \\ &= -2\beta\lambda\int_\sigma^\infty dx\cos(kx)w_{tail}(x)\end{aligned} \tag{40}$$

where we have made explicit the one-dimensional Fourier transform. The integral is easily performed analytically. The mean-field result for the structure factor is

$$S_{MF\lambda}(k) = [1 - \rho \hat{c}^{(2)}_{MF\lambda}(k)]^{-1} . \tag{41}$$

It is straightforward to show that the compressibility sum rule $S(0) = \rho \beta^{-1} \chi$ is satisfied by $S_{MF\lambda}(0)$ with the pressure $P_{MF\lambda}$ given by (35). This means that an unphysical feature in the compressibility will be directly reflected in $S_{MF\lambda}(0)$. We shall see that other unphysical features may also arise in $S_{MF\lambda}(k)$.

Results for the structure factor, for size-ratio 0.9, are shown in Figs. (8-10) for fugacities $z_p\sigma$ =3, 5 and 9, respectively. The blue solid lines refer to the exact results for the full potential, i.e. for $\lambda$=0. In the case $z_p\sigma$ =3 and $\rho\sigma$ = 0.2, Fig.8a, treating 10% of the tail potential in mean-field ($\lambda$=0.1) yields an excellent approximation to the exact result. The standard RPA that treats all the tail in mean field ($\lambda$=1) gives a poor result; $S_{MF\lambda}(0)$ lies well below the exact result and the amplitude of the oscillations is underestimated. The results for a higher density $\rho\sigma$ = 0.5 are displayed in Fig.(8b) where we plot the *inverse* of the structure factor. Once again the results for $\lambda$=0.1 agree very well with the exact results but for $\lambda$=1, $[S_{MF\lambda}(k)]^{-1}$ takes negative values for $k\sigma \leq 5$ implying an unphysical divergence of the structure factor at non-zero wavenumbers. In Fig.9 the corresponding plots are made for $z_p\sigma$ =5 for the same two densities. For $\lambda$=0.1 the mean-field treatment performs very well at both densities. However, for $\lambda$=1 it is very poor; for the higher density, Fig.9(b), there is a region near $k\sigma$=5 where $[S_{MF\lambda}(k)]^{-1}$ is negative.

The most striking results are those for the most attractive pair potential $z_p\sigma$ =9, shown in Fig.10. For $\rho\sigma$ = 0.02, a very dilute gas, the exact structure factor exhibits very pronounced oscillations with a wavelength close to $\sigma$-see Fig.10(a). These are captured well by the mean-field treatment with $\lambda$=0.1. But for $\lambda$=1 (standard RPA) the structure factor is close to that of an ideal gas. There are only very weak oscillations and $S_{MF\lambda}(0)$ is only slightly greater than unity. The fluid has a large compressibility at this (small) density which is not accounted for by the standard RPA, as can be ascertained from Fig.7. Given that the fluid is highly compressible it is not too surprising that the structure factor (the density-density response function) exhibits such pronounced oscillations. For the higher density $\rho\sigma$ = 0.4 the exact value of $S(0)$ is about 9 and the oscillations in $[S(k)]^{-1}$ shown in Fig. 10(b) are extreme; the height of the peak in $S(k)$ at $k\sigma$=2π is about equal to $S(0)$. The mean-field treatment with $\lambda$=0.1 captures well the main features but $[S_{MF\lambda}(k)]^{-1}$ is negative near $k$=0 reflecting the fact that $(\partial P_{MF\lambda}/\partial \rho)_{z_p}$ is negative for this density –see Fig.7(c). For $\lambda$=1 the (standard RPA) structure factor is very poor; the oscillations are not well-accounted for and $[S_{MF\lambda}(k)]^{-1}$ is negative near $k\sigma$=0 and $k\sigma$=5.

We emphasize that accounting for the details of the structure factor for this particular model fluid at large $z_p\sigma$ is a challenging test for any theory. The results for the one-dimensional model show that including a significant part (although not all) of the attraction in the reference system can lead to much more accurate results for the equation of state and the structure factor than treating the full attraction via the RPA. This demonstrates that splitting potentials in ways different from the traditional divisions and then treating just a part of the attractive tail using the LMF theory might be very fruitful.

**V DISCUSSION**

In this paper we have re-visited the basis of LMF theory for simple fluids. We showed in Sec.II that the LMF equation (15), as introduced by Weeks et.al.[6], can be readily derived using an argument based on the one-body direct correlation function and that LMF is therefore firmly embedded in standard DFT treatments of attractive interatomic forces. For interatomic potentials, or inter-particle potentials in colloidal fluids, possessing a hard-core the natural reference system is the hard sphere fluid and for such a system modern DFT theories remain accurate up to the freezing density [1-3]. Practitioners would argue that DFT describes the non-uniform hard-sphere fluid with accuracy comparable to that of simulation. It follows that for model fluids with hard-core repulsion LMF is equivalent to the standard (mean-field) treatment of the attractive tail, commonly employed in DFT calculations[3,4] .For such models LMF, with a simulation treatment of the reference fluid, will not improve significantly upon DFT. In this context we refer to the comments on DFT made by Weeks in his admirable review article[9] in 2002. Weeks' Eq.(29) for a hard-sphere reference fluid is identical to our (19).However, in contrast to our present argument, Weeks takes the view that finding suitable functionals $F_{exR}$ is difficult, even for hard-spheres and he appears not to advocate solving (19). We agree with Weeks that LMF does have certain advantages over DFT when one considers fluids where the interatomic repulsion is soft. By making a suitable division of the pair potential, e.g. that of WCA, one can treat a soft repulsive, non-uniform, reference fluid using simulation. DFT approaches for soft repulsive, e.g. inverse power law potentials, are less well-developed than for hard-core models but there are promising developments [3, 34].

Within the DFT framework there are attempts to improve upon the mean-field treatment of attraction. In essence these seek to provide functionals that generate a pair direct correlation function for the uniform fluid, $c^{(2)}(r;[w])$, more accurate than the RPA (16). Although there is progress [3, 35], a systematic and tractable scheme for improving upon the RPA is arguably still missing. The obvious route to improvement is to start with the formally exact result [4] for the difference in free energy functionals on the r.h.s. of Eq.(13)

$$F_{ex}[\rho;[w]] - F_{exR}[\rho;[u_0]] = \frac{1}{2}\int_0^1 d\lambda \int d\mathbf{r} \int d\mathbf{r}' \rho^{(2)}(\mathbf{r},\mathbf{r}';u_\lambda) u_1(|\mathbf{r}-\mathbf{r}'|) \qquad (42)$$

where $\rho^{(2)}(\mathbf{r},\mathbf{r}';u_\lambda)$ is the pair distribution for a fluid with pair potential $u_\lambda(r) = u_0(r) + \lambda u_1(r)$ and the coupling parameter $\lambda$ is varied from 0 to 1. The RPA (and the LMF), c.f. Eq.(14), follows from the mean-field approximation

$$\int_0^1 d\lambda \rho^{(2)}(\mathbf{r},\mathbf{r}';u_\lambda) \approx \rho(\mathbf{r})\rho(\mathbf{r}'). \qquad (43)$$

One can attempt to include correlations by approximating the r.h.s. of (43) by say $\rho(\mathbf{r})\rho(\mathbf{r}')g(|\mathbf{r}-\mathbf{r}'|;u_0)$. However, one must then specify the density at which $g$, the radial distribution of the uniform reference fluid, is evaluated. Generally this leads to a loss of thermodynamic consistency -one of the virtues of DFT that one does not wish to abandon-see below. Nevertheless it is important to pursue such approaches as these can bring new insight to improving upon the standard mean-field treatment.

In Sec.III we examined drying at a hard wall within the framework of LMF employing our DFT perspective. We showed that the effective reference potential $V_{eff}(z)$ must exhibit a rich structure that can include damped oscillations at large distances as well as the expected repulsive hump required to form the drying layer of gas-see Figs. 2,3,5,6.

We were able to extract such detailed information because we used DFT; our approach is completely equivalent to solving (19) for the hard-sphere reference fluid. It is a challenge to obtain the same level of detail via simulations. One might also enquire whether the LMF approach used in Ref.[6], a study of the LJ liquid on a near critical isotherm adsorbed at a hard-wall, could provide similar detail for drying situations. Of course, drying at a planar wall is a particular example of solvation. The same DFT that we employ here has been used to elucidate some of the subtle non-analyticities of the interfacial free-energy and adsorption associated with drying at very large spherical solutes for both short-ranged [36] and power-law (dispersion) [37] interatomic forces. In the light of our present analysis one *could* say that those studies of solvation, and more generally the huge number of DFT studies of adsorption and interfacial phenomena based on (17), are LMF calculations using an accurate DFT for the hard-sphere reference fluid. Katsov and Weeks [8] used LMF to investigate the density profile and solvation free energy of a LJ liquid adsorbed at a hard-sphere solute of increasing radius, comparing their results with those from earlier simulations [38]. In Ref. [8] a modification was made to the LMF equation (15) and a generalized linear response theory [7] (*not* simulation) was used to determine the structure of the non-uniform, hard-sphere reference fluid. Although there are differences it is clear that the existing DFT and LMF approaches to solvation have much in common and it would be interesting to make detailed comparisons. In particular one would like to know whether the generalized linear response treatment for the reference fluid captures the subtle features, arising from drying, found for very large hard-sphere solutes [36, 37]. DFT results [39] for radii up to $6\sigma$ show good qualitative agreement with the simulation data [38] but careful comparisons should be made between the theories and simulation at the same value of $\delta\mu$, or the same ratio of bulk to coexisting liquid density, for a wide range of radii.

An important feature of the DFT that we employ is that the density profiles and free energies it generates via minimization of the grand potential functional (17) automatically satisfy the sum rules that link thermodynamic properties to structure (density profiles or correlation functions)[4]. Some DFT treatments and most integral equation approaches for non-uniform fluids do not satisfy this requirement. These have severe disadvantages. In particular determining the location of interfacial phase transitions, such as wetting, pre-wetting and capillary condensation, is problematical. More generally one asks: Which route does one choose to calculate the free energy (grand potential) of the non-uniform fluid? This is also an issue for LMF treatments that are not based, like ours, in DFT. As emphasized by Katsov and Weeks [8] LMF theory focuses on liquid structure as do most integral equation methods. They also point out that different routes to the same thermodynamic property can give different answers but they opt for a 'virial route' to the calculation of the solvation free energy of the hard-sphere solute that adopts scaled particle methodology and requires only the density at contact. In the standard DFT approach, for better or worse, there is only one route to thermodynamics. It would be useful to establish how well or poorly (non-DFT based) LMF treatments perform when it comes to satisfying the Gibbs adsorption equation and the contact theorem for the fluid at a planar or spherical hard-wall.

In keeping with Weeks and co-workers we have emphasized that the efficacy of LMF lies in making an intelligent division of the interatomic pair potential of the full system, $w(r)$, into a reference potential and a remainder that one might treat in mean-field. In Sec.IV we focused on different ways of implementing this division for the case of a uniform fluid. For the demanding case of a one-dimensional fluid with a linear attractive potential (29) we investigated how both the equation of state and the structure factor depend on the choice of reference system. The message that emerges is that traditional divisions might not be the most effective and that one might consider incorporating some part of the attraction into the reference fluid. Similar conclusions emerge from the paper of Denesyuk and Weeks[40] which is concerned with the application of a simplified version of LMF to a uniform ionic mixture. The authors argue for a reference (mimic) system which includes short-ranged parts of the Coulomb interactions. The smoothing length $\Gamma$, introduced at the beginning of our Sec.IV, is chosen so that the remainder of the pair potentials are sufficiently slowly varying to be treated in the LMF approach. For an accurate treatment this requires simulations of the reference system - there is no suitable DFT or other approximation. In addition Ref.[40] investigates the structure and electrostatic energy of the mixture (see Sec.V) using a RPA for the charge structure factor that is equivalent to the one we implement in Sec.IV for our one-dimensional model. For the particular division (choice of $\Gamma$) that is implemented the structure factors are very close to the 'exact' simulation results for the full pair potentials. Ref.[40] also mentions that LMF uses a mean-field average of long-ranged interactions similar in spirit to the RPA used in DFT but does not make explicit the connection between the two approaches- as we do in the present paper. We are grateful to a referee for bringing Ref.[40] to our attention.

**ACKNOWLEDGEMENTS**

The authors were motivated to study LMF as a result of listening to stimulating talks by John Weeks and by Jocelyn Rodgers. A.J.A. acknowledges support from the European Union under Grant No. PITN-GA-2008-214919 (MULTIFLOW) and R.E. from the Leverhulme Trust.

**Figure Captions**

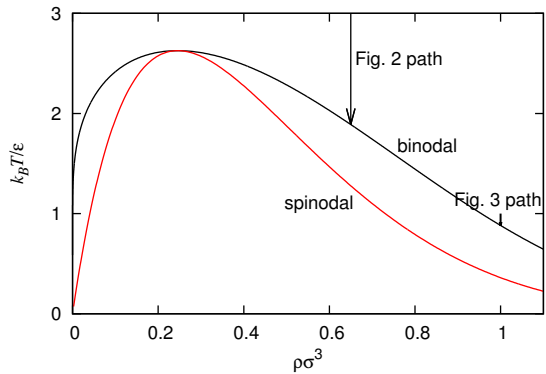

FIG.1. Phase diagram of the bulk hard-core Yukawa fluid with $Z$=1, calculated using the RPA treatment of the attractive tail. The vertical arrows denote two paths of constant density followed in the drying studies of Figs. 2 and 3.

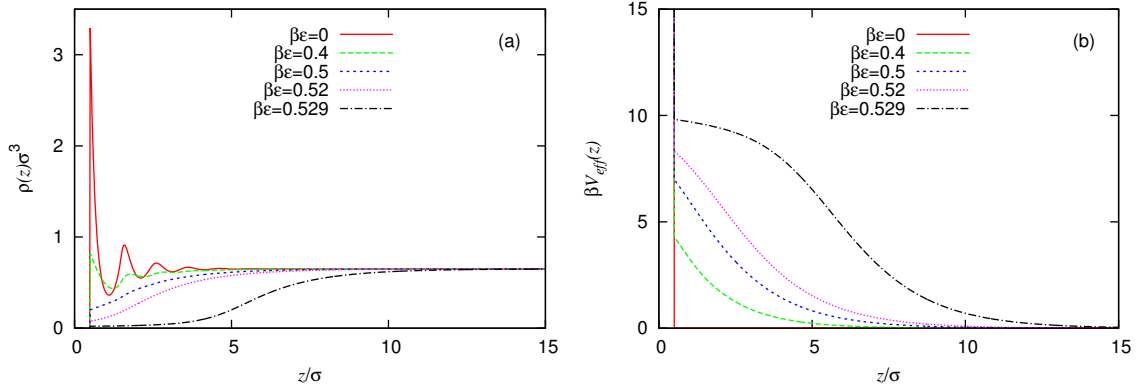

FIG.2. (a) Density profiles $\rho(z)$ for the hard-core Yukawa fluid with $Z=1$ and constant bulk density $\rho\sigma^3=0.65$ adsorbed at a hard–wall, located at $z=\sigma/2$. As $\beta\varepsilon$ is increased, equivalent to decreasing the temperature along the first path in Fig.1, we observe the erosion of oscillations and the growth of a drying film of gas. For this bulk density the binodal is at $\beta\varepsilon=0.5293$. The result for $\beta\varepsilon=0$ corresponds to the hard-sphere fluid. (b) The corresponding effective external potentials $V_{eff}(z)$ that would have to be exerted by a wall on the hard-sphere fluid in order to have identical density profiles. The effective potentials decay monotonically (exponentially) to zero as $z \to \infty$ and $\rho(z) \to \rho$.

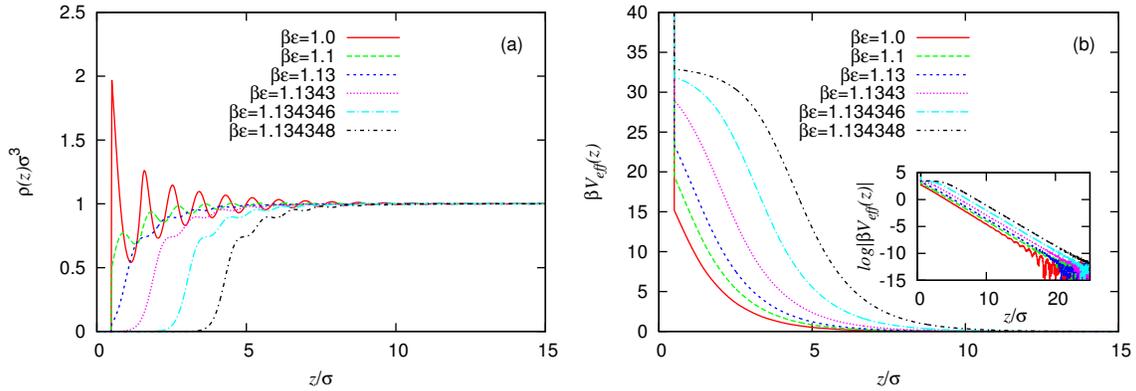

FIG.3. (a) As in Fig.2(a) but now for the path at constant bulk density $\rho\sigma^3=1.0$. We observe the growth of the drying film on approaching the binodal located at $\beta\varepsilon=1.134349$. Note that the density profile for the thickest film shown, with $\beta\varepsilon=1.134348$, remains weakly oscillatory. (b) The corresponding effective external potentials $V_{eff}(z)$ that would have to be exerted by a wall on the hard-sphere fluid in order to have identical density profiles. The inset shows the logarithm of the effective potentials in the tail region-see text.

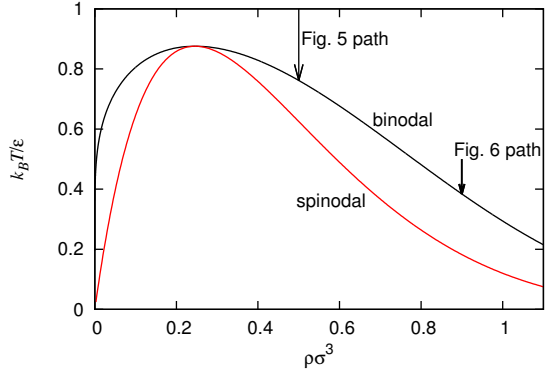

FIG.4. Phase diagram of the bulk hard-core Yukawa fluid with $Z=3$, calculated using the RPA treatment of the attractive tail. The vertical arrows denote two paths of constant density followed in the drying studies of Figs. 5 and 6.

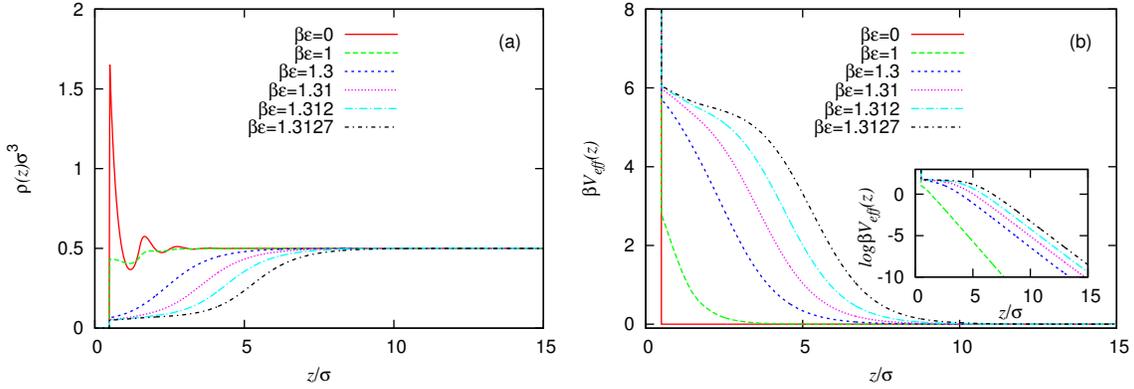

FIG.5. (a) Density profiles $\rho(z)$ for the hard-core Yukawa fluid with $Z=3$ and constant bulk density $\rho\sigma^3=0.5$ adsorbed at a hard –wall, located at $z=\sigma/2$. As $\beta\varepsilon$ is increased, equivalent to decreasing the temperature along the first path in Fig.4, we observe the erosion of oscillations and the growth of a drying film of gas. For this bulk density the binodal is at $\beta\varepsilon=1.31279$. The result for $\beta\varepsilon=0$ corresponds to the hard-sphere fluid. (b) The corresponding effective external potentials $V_{eff}(z)$ that would have to be exerted by a wall on the hard-sphere fluid in order to have identical density profiles. The inset shows the logarithm of the effective potentials in the tail region; the potentials decay monotonically (exponentially) to zero as $z\to\infty$ and $\rho(z)\to\rho$.

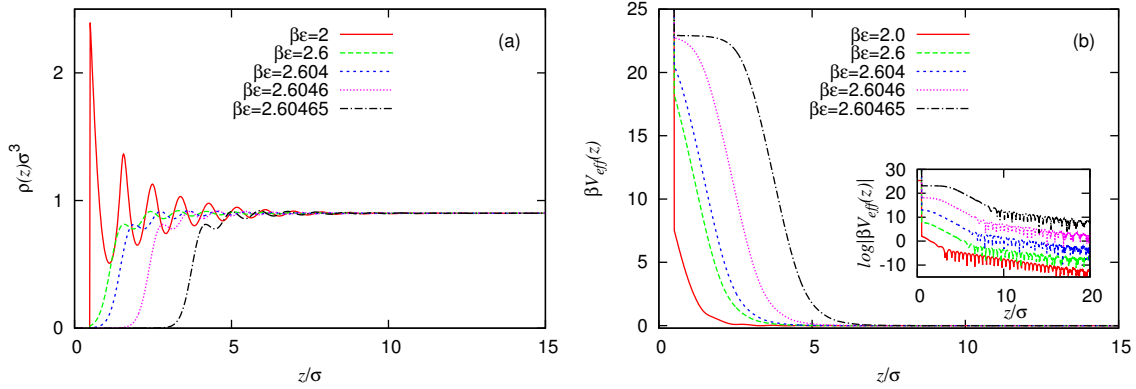

FIG.6. (a) As in Fig.5(a) but now for the path at constant bulk density $\rho\sigma^3=0.9$. We observe the growth of the drying film on approaching the binodal located at $\beta\varepsilon=2.604651$. Note that for the thickest film shown, with $\beta\varepsilon=2.60465$, the density profile exhibits damped oscillations at large $z$. (b) The corresponding effective external potentials $V_{eff}(z)$ that would have to be exerted by a wall on the hard-sphere fluid in order to have identical density profiles. The inset shows the logarithm of the effective potentials in the tail region. These have been shifted vertically for clarity. The potentials exhibit exponentially damped oscillations as $z\to\infty$ and $\rho(z)\to\rho$ -see text.

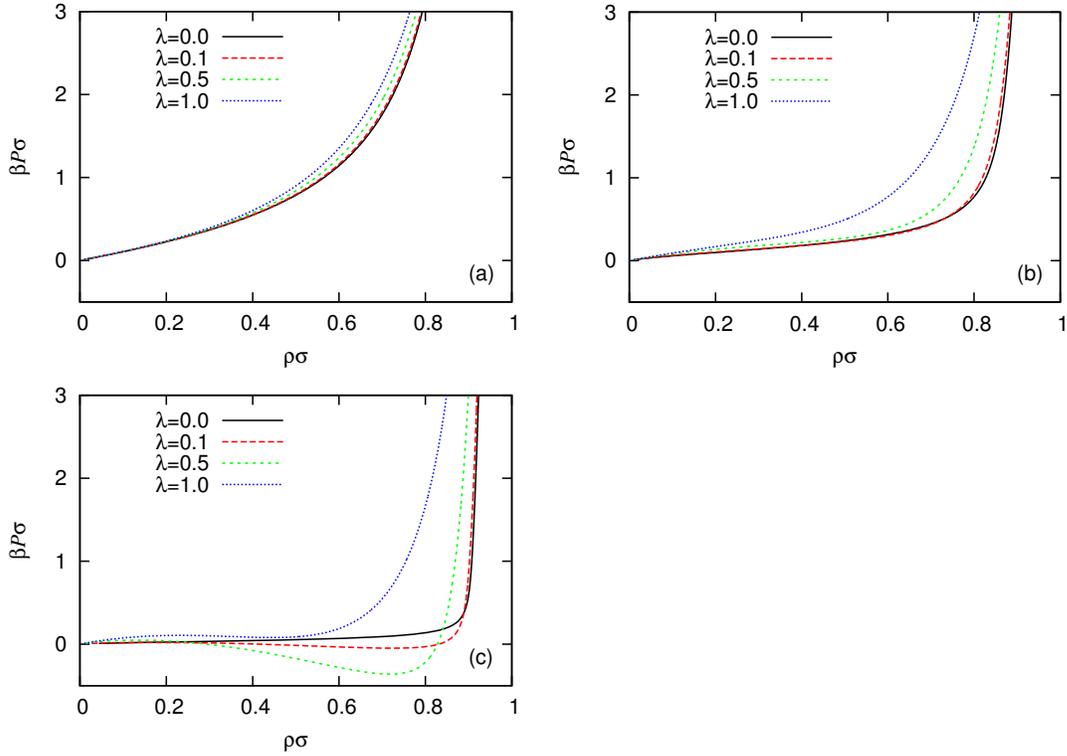

FIG.7. Pressure $P$ as a function of density $\rho$ for three different polymer fugacities: (a) $z_p\sigma =1$, (b) $z_p\sigma =5$ and (c) $z_p\sigma =9$, for the one-dimensional model fluid described by the pair potential (28, 29) and size-ratio $\sigma_p/\sigma=0.9$. The blue solid lines, for $\lambda=0$, are the exact results. The other curves treat increasing amounts of the attractive tail potential, $\lambda w_{tail}(x)$, within mean field approximation. $\lambda=1$ corresponds to the standard RPA for the full tail. Note that for $z_p\sigma =9$ unphysical van der Waals loops develop even for $\lambda=0.1$.

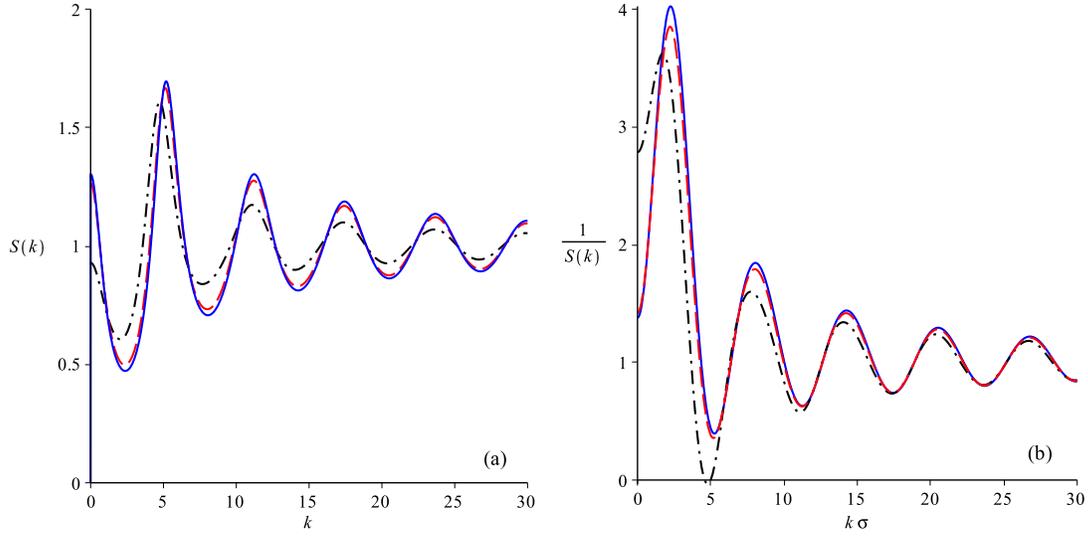

FIG.8. Structure factor $S(k)$ of the one-dimensional model fluid for size-ratio $\sigma_p/\sigma=0.9$ and $z_p\sigma=3$. The blue solid lines, for $\lambda=0$, are the exact results. The other two curves treat increasing amounts of the attractive tail potential, $\lambda w_{tail}(x)$, within mean-field approximation: $\lambda=0.1$ (red dashed) and $\lambda=1$ (black dash-dot). (a) $S(k)$ for $\rho\sigma=0.2$. (b) The *inverse* of $S(k)$ for $\rho\sigma=0.5$. Note that for $\lambda=1$ (standard RPA) this function takes negative values near $k\sigma=5$ corresponding to an unphysical divergence of $S(k)$.

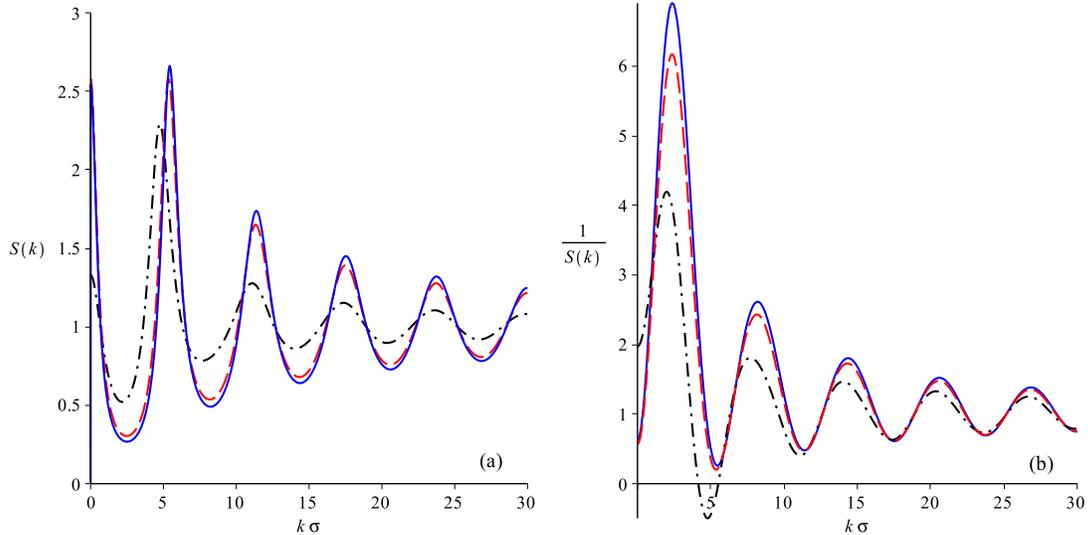

FIG.9. As in Fig.(8) but now for $z_p\sigma=5$. $\lambda=0$ (blue solid), $\lambda=0.1$ (red dashed) and $\lambda=1$ (black dash-dot).(a) $S(k)$ for $\rho\sigma=0.2$. (b) The *inverse* of $S(k)$ for $\rho\sigma=0.5$. Note that for $\lambda=1$ (standard RPA) this function takes negative values near $k\sigma=5$.

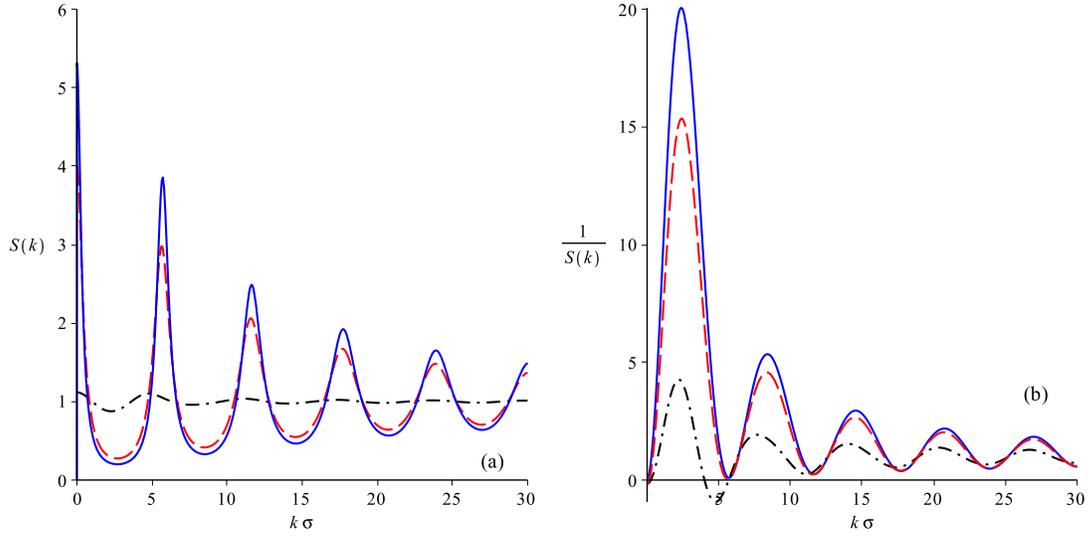

FIG.10. As in Fig.(8) but now for $z_p\sigma =9$. $\lambda=0$ (blue solid), $\lambda=0.1$ (red dashed) and $\lambda=1$ (black dash-dot).(a) $S(k)$ for $\rho\sigma=0.02$. Note that for $\lambda=1$ (standard RPA) $S(k)$ is close to that for an ideal gas. (b) The *inverse* of $S(k)$ for $\rho\sigma=0.4$. Note that for $\lambda=0.1$ this function is negative near $k=0$. The density $\rho\sigma=0.4$ lies inside the spinodal region in Fig.7(c). For $\lambda=1$ the function is negative near $k\sigma=0$ and $k\sigma=5$.